\documentclass[aps,pr,floatfix,superscriptaddress,twocolumn,showpacs]{revtex4}
\usepackage{epsfig}
\usepackage{amsmath}
\usepackage{graphicx,psfrag}
\usepackage{dcolumn}
\usepackage{bm}
\usepackage{slashed}
\usepackage{xcolor}
\usepackage[makeroom]{cancel}

\usepackage[title]{appendix}

\definecolor{darkgreen}{rgb}{0,0.7,0.3}

\newcommand{\beq}{\begin{equation}}
\newcommand{\eeq}{\end{equation}}
\newcommand{\bea}{\begin{eqnarray}}
\newcommand{\eea}{\end{eqnarray}}
\newcommand{\hf} {\frac{1}{2}}

\newcommand{\nn}{\nonumber\\}

\newcommand\fig[1]     {Fig.\,{\ref{#1}}}

\newcommand\app[1]     {Appendix~\ref{#1}}

\def\Tr{{\rm Tr}}

\def\eq#1{(\ref{#1})}
\def\s0#1#2{\mbox{\small{$ \frac{#1}{#2} $}}}
\def\0#1#2{\frac{#1}{#2}}

\def\mr#1{{\mathrm{#1}}}

\begin{document}

\title{Pseudo--Periodic Natural Higgs Inflation}

\author{I. G. M\'ari\'an}
\affiliation{University of Debrecen, P.O.Box 105, H-4010 Debrecen, Hungary}

\author{N. Defenu}
\affiliation{Institut f\"ur Theoretische Physik, Universit\"at Heidelberg, D-69120 Heidelberg, Germany}

\author{U. D. Jentschura}
\affiliation{Department of Physics, Missouri University of Science and Technology, Rolla, Missouri 65409-0640, USA}
\affiliation{MTA--DE Particle Physics Research Group, P.O.Box 51, H--4001 Debrecen, Hungary}
\affiliation{MTA Atomki, P.O. Box 51, H--4001 Debrecen, Hungary} 

\author{A. Trombettoni}
\affiliation{CNR--IOM DEMOCRITOS Simulation Center, Via Bonomea 265, I-34136 Trieste, Italy}
\affiliation{SISSA and INFN, Sezione di Trieste, Via Bonomea 265, I-34136 Trieste, Italy}

\author{I. N\'andori}
\affiliation{University of Debrecen, P.O.Box 105, H-4010 Debrecen, Hungary}
\affiliation{MTA--DE Particle Physics Research Group, P.O.Box 51, H--4001 Debrecen, Hungary}
\affiliation{MTA Atomki, P.O. Box 51, H--4001 Debrecen, Hungary}

\begin{abstract} 
Inflationary cosmology represents a well-studied framework to describe the expansion of space 
in the early universe, as it explains the origin of the large-scale structure of the cosmos and the 
isotropy of the cosmic microwave background radiation. The recent detection of the Higgs boson 
renewed research activities based on the assumption that the inflaton could be identified with the 
Higgs field.  At the same time, the question whether the inflationary potential can be be extended 
to the electroweak scale and whether it should be necessarily chosen {\em ad hoc} in order to be 
physically acceptable are at the center of an intense debate. 
Here, we perform the slow-roll analysis of the so-called Massive Natural Inflation (MNI) model 
which has three adjustable parameters, the explicit mass term, a Fourier amplitude $u$, and a 
frequency parameter $\beta$,
in addition to a constant term of the potential. This theory has the advantage 
to present a structure of infinite non-degenerate minima and is amenable to an easy integration 
of high-energy modes. We show that, using PLANCK data, one can fix, in the large $\beta$-region,
the parameters of the model in a unique way.
%%% and reduce the number of free parameters to zero.
We also demonstrate that the value for the parameters chosen at the cosmological scale does not 
influence the results at the  electroweak scale. We argue that other models can have similar 
properties both at cosmological and electroweak scales, but with the MNI model one can complete 
the theory towards low energies and easily perform  
the integration of modes up to the electroweak scale, producing the correct order-of-magnitude for 
the Higgs mass. 
\end{abstract}

\pacs{98.80.-k,14.80.Cp,11.10.Hi}

\maketitle 

\tableofcontents

%------------------------------------------------------------------------------------ 
\section{Introduction}
\label{sec_intro}
%------------------------------------------------------------------------------------

Exponential expansion of the early universe can be explained by cosmic inflation, 
a theory which is developed to explain major issues such as the origin of the large-scale 
structure, the flatness of the universe, the horizon problem, the absence of monopoles and 
in general properties of Cosmic Microwave Background Radiation (CMBR) \cite{inflation}. 
A very comprehensively studied work hypothesis is that a hypothetical scalar field, i.e., the 
inflaton particle, is responsible for inflation which is caused by the slow-roll motion starting 
from a metastable false vacuum towards the real vacuum \cite{linde,albrecht_steinhardt}. 
However, it is still a matter of discussion whether a particle physics mechanism can be 
associated to inflation or in general whether we have or not a reliable approach to inflation 
\cite{criticism_1,criticism_2}.

The recent detection of the Higgs boson renewed research activity where the inflaton is 
identified with the Higgs field. Indeed, it seems to be possible to extrapolate the Standard 
Model (SM) of particle physics up to very high energies, and the most "economical" choice 
would be to use the same scalar field in order to describe the Higgs and inflationary physics. 
Nevertheless, various issues related to Higgs-inflation, like stability of the Higgs potential 
or the exit from the inflationary phase,  need to be explained. As an example, it was argued 
in Ref.~\cite{bezrukov_rubio_shapo} that the traditional Higgs inflation can be possible within a 
minimalistic framework even if the SM vacuum is not completely stable. Furthermore, the 
importance of renormalization group (RG) running on the stability question is highlighted
(see, e.g., Refs.~\cite{rg_and_stability,higgs_frg_1,higgs_frg_2}). One can also mention the 
possible relation of the stability problem to unknown new physics, as in  
Refs.~\cite{new_physics_stablity_1,new_physics_stablity_2}, where the analysis is 
performed in a flat as well as curved spacetime background, showing that new physics 
can have a significant impact on the stability condition of the vacuum.

There are, however, several drawbacks of Higgs-inflation and in general cosmological 
inflation, which require further studies and explanation. A serious problem related to the 
identification of the Higgs with the inflaton is that a reliable model should work at 
cosmological as well as electroweak energy scales. Thus, a single model should be used 
to explain, simultaneously, recent data on CMBR thermal fluctuations and those measured 
at the electroweak scale.  Another issue is, for example, the choice for the inflationary 
potential which means that there is a plethora of scalar models available in the literature 
\cite{encyc} which all work well at cosmological scales and any choice of them seems to be 
{\em ad hoc}. Under the assumption of a relation of Higgs and inflationary physics, the 
solution for the former problem can automatically represent a solution for the latter, i.e., 
the requirement for a scalar potential viable both for inflationary and Higgs physics 
drastically reduces the number of admissible proposals. Indeed, a mode integration 
treatment can help to relate parameters of a candidate model at both energy scales and 
it is expected to give us a tool to reduce the number of viable scenarios for inflation. 
 
In order to gain insights on the previous point, an important qualitative issue is provided 
by the task of understanding the structure of the inflaton potential. A simple harmonic 
potential $\phi^2$ has a single minimum which is known as the quadratic large-field 
inflationary (LFI) potential. One can add more minima through higher-order powers of the 
form $\phi^{2n}$.  A general question is whether and how much one has to deviate from the 
$\phi^2$-Gaussian form. In Ref.~\cite{periodic_inflaton_test}, the ``$\phi^2$ or not-$\phi^2$'' 
issue was tested on the simplest inflationary potential: constraints were obtained and the 
relevance of non-Gaussianity was discussed. From the opposite point of view, a very much 
``not-$\phi^2$'' potential is the one having infinitely many minima, at the same energy. 
In this logic, one can explore a periodic potential of a form having infinitely many minima, 
$V_{\rm NI}(\phi) = u\left[\cos(\beta \phi)-1\right]$ which is known as the
Natural Inflation (NI) or pseudo-Nambu-Goldstone boson model, 
while in field theory and condensed matter,
it is denoted as the sine-Gordon model~\cite{coleman}. 
It has also been proposed and studied as a viable inflationary scenario 
\cite{periodic_inflaton_1,periodic_inflaton_2,periodic_inflaton_test_2_1,periodic_inflaton_test_2_2,%
periodic_inflaton_test_2_3,periodic_inflaton_test_2_4,periodic_inflaton_test_2_5,periodic_inflaton_test_2_6,%
periodic_inflaton_test_2_7,periodic_inflaton_test_2_8,periodic_higgs} and 
to construct a convenient scalar sector by incorporating the periodic scalar axion potential 
too~\cite{periodic_higgs}. It was shown that the NI potential is able to produce agreement 
with PLANCK results~\cite{planck_1,planck_2,planck_3} on the thermal fluctuations of the 
cosmic microwave background radiation (CMBR) with a better agreement than the simplest 
LFI potential~\cite{periodic_inflaton_test} and that in $d=4$, it has a 
single phase~\cite{periodic_higgs}. 

In this paper, we propose and use a relatively simple scalar model which has the advantage 
of having an overall $\phi^2$ form shifted by a constant and retaining however a structure 
with infinitely many minima. We show that the proposed potential {\it (i)} is a viable choice 
for inflation in its large $\beta$ limit, and {\it (ii)} serves as a possible extension of the SM Higgs 
potential. The construction of the potential is based on the so-called 
massive sine-Gordon model where a term sinusoidal in the field is added to the standard 
quadratic mass term; this potential has already received significant attention in statistical field 
theory \cite{msg_lpa_1,msg_lpa_2,msg_lpa_3,msg_beyond_lpa}. We denote our proposal
as the Massive Natural Inflation (MNI) model,
\begin{equation}
\label{MNI}
V_{\rm MNI}(\phi) =   V_0 + \hf m^2 \phi^2 + u \cos(\beta \phi), 
\end{equation}
where $m$ is an explicit mass term, $u$ is the Fourier amplitude, 
$\beta$ is the frequency, and $V_0$ is a constant (field-independent) term which is 
either chosen to be equal to zero or $-u$ or adjusted 
in such a way that the global minimum value of the potential is retained
at $V_{\rm min} = V_{\rm MNI}(\phi_{\rm min}) = 0$.
In general, the constant term can also be considered as a free parameter, however, we will 
demonstrate that it does not modify the slow-roll analysis (comparison with PLANCK data) in 
the large $\beta$ limit and plays no role in the mode integration. The 
MNI potential (\ref{MNI}) has, indeed, an infinite number of minima, separated by an amount 
of energy depending on the ratio of the coefficients multiplying the two terms, and ranging 
from the limit of infinite degenerate minima to a single non-degenerate absolute minimum.

Therefore, the MNI model appears to be an excellent candidate for cosmological inflation 
and a viable extension for the SM Higgs potential. By mode integration, in the following, we 
show that it is possible to relate the parameters of the model at various energy scales 
(cosmological and electroweak). The next question is whether this constitutes a unique feature 
of the MNI model. Alternatively, one might ask if one can find, in general, other scalar models 
(with at least two parameters) which have the same properties. As a partial answer to this
question, we observe that, due to the fact that the studied model has two energy scales 
(cosmological and electroweak ones), which are far from each other, it is expected that any suitable 
theory in the low-energy (IR) limit (defined here as the electroweak scale) becomes insensitive 
to its high-energy (UV) behavior (here understood as the inflationary scale). This is demonstrated 
by us using the above mentioned MNI model, and we argue that this is a property shared by a 
number of related models. We will come back to these points in our conclusions.

This paper is organized as follows: In Section \ref{cs_sec} we discuss the application of the 
MNI potential at cosmological scale, considering the comparison with PLANCK data and 
studying in detail the limit of high-frequency (high $\beta$). In Section \ref{es_sec}, we consider 
the theory at electroweak scale, presenting a discussion of the relation between the 
Higgs mass and the parameters of the MNI model at low energies. The connection between the 
cosmological and the electroweak scales using the MNI model is detailed in Section \ref{mi_sec}, 
where we show explicitly that at low energy, the theory exhibits UV-insensitivity. The latter is clearly 
demonstrated by the derivation presented there, where the UV mass cancels out. Our conclusions 
are finally presented in Section \ref{concl}.

%------------------------------------------------------------------------------------
\section{Cosmological Scale}
\label{cs_sec}
%------------------------------------------------------------------------------------

Early universe and Higgs physics are examples where scalar fields find a natural role to play in 
standard models of cosmology and particle physics. Since scalar fields can mimic the equation 
of state required for exponential expansion of the early universe, various types of scalar potentials 
have been proposed in inflationary cosmology. The simplest of these scenarios is provided by 
the slow-roll single-field models with minimal kinetic terms \cite{encyc}. A good candidate for an 
inflationary potential should have a small number of free parameters which serves as the first 
condition for a reliable model, and the primary example is the well studied quadratic, LFI potential 
having the form $V = \tfrac12 \, m^2 \phi^2$.   

Inflationary models should be as well tested on whether (and how well) they can reproduce the 
observed data of various experiments such as the PLANCK mission \cite{planck_2,planck_1,planck_3}, 
which measures thermal fluctuations of cosmic microwave background radiation (CMBR). This serves 
as a minimal requirement to obtain a viable scenario for the post-inflationary period. 

We consider, motivated by the reasons exposed in Section \ref{sec_intro}, the MNI model (\ref{MNI}). 
With the aim of performing checks on our results, we also consider three variants of the MNI model 
(\ref{MNI}) provided by 
\begin{subequations}
\label{VVV}
\begin{align}
\label{V1}
V_1(\phi) =& \; \hf m^2 \phi^2 + u \cos(\beta \phi)\,, 
\\[0.1133ex]
\label{V2}
V_2(\phi) =& \; \hf m^2 \phi^2 + u\left[\cos(\beta \phi)-1\right]\,,
\\[0.1133ex]
\label{V3}
V_3(\phi) =& \; \hf m^2 \phi^2 + u\cos(\beta \phi) - V_{\rm min}\,,
\end{align}
\end{subequations}
where $V_{\rm min}$ is introduced as an adjustable parameter 
to keep the  global minimum of the potential at zero.

%------------------------------------------------------------------------------------
\subsection{Slow-Roll Analysis}
\label{slow_roll_sec}
%------------------------------------------------------------------------------------

Since we use three different values for the field independent constant, the MNI model (and its considered 
variants), in addition to the usual normalization, have two adjustable parameters, the ratio $u/m^2$ and 
the frequency $\beta$. The remaining two adjustable parameters can be fixed at cosmological scales by 
the standard slow-roll analysis where one has to calculate the following parameters
(see e.g., Ref.~\cite{MP}),
\beq
\epsilon\equiv \hf m_p^2 \left(\frac{\displaystyle V'}{\displaystyle V}\right)^2, \qquad
\eta\equiv m_p^2 \frac{\displaystyle V''}{\displaystyle V} \,,
\eeq
with the Planck mass $m_p^2 = 1/(8\pi G)$. Inflation is in progress if the conditions $\epsilon \ll 1$ 
and $\eta \ll 1$, are fulfilled. If one of these parameters assumes a value on the order of unity, 
then inflation stops. In order to 
have a prolonged inflation, the E-fold number $N$ which is defined by 
\begin{equation}
N\equiv \dfrac{1}{m_p^2} \int_{\phi_i}^{\phi_f} {\rm d}\phi \, \frac{V}{V'}
\label{int}
\end{equation} 
should be in the range $50 < N < 60$ where $\phi_i$ and $\phi_f$ are the initial and final values
of the vacuum expectation value (VeV) of the field over inflation.

In a $(d=4)$-dimensional field theory, the physical dimension of $V$ has to be 
equal to mass$^4$, so one can introduce dimensionless 
variables accordingly. By using an arbitrary scale $k$ which has the dimension 
of mass, one has 
\begin{align}
\label{dimless_param}
\tilde{V} = & \; V/k^4, \quad \tilde{\phi} = \phi/k , \quad
\tilde{u} = u/k^4 \,, 
\nonumber\\[0.1133ex]
\tilde{m} = & \; m/k, \quad \tilde{\beta} = k \beta \,: 
\end{align}
In order to obtain, from PLANCK data, 
the parameters of the MNI model (\ref{MNI}), 
we will work in reduced Planck units $c \equiv \hbar \equiv1$ and 
$m_p^2 \equiv 1$, corresponding to choose $k=m_p$ 
in Eqs. (\ref{dimless_param}). 
As routinely done in literature, see e.g. Ref.~\cite{MP}, we also introduce 
the scalar tilt $n_s$ and tensor-to-scalar ratio $r$. They 
are related to the slow-roll 
parameters by the two relations 
$n_s \approx 2 \eta -6 \epsilon+1$ and $r\approx 16 \epsilon$ 
\cite{MP}. 
From the conditions $\epsilon(\phi_{f,\epsilon})=1$ and $\eta(\phi_{f,\eta})=1$, the final value of
the field VeV can be calculated in Eq.~(\ref{int}) as 
$\phi_f \equiv \max{\left( \phi_{f,\epsilon},\phi_{f,\eta} \right)}$.
Knowing $\phi_f $, one can determine its initial value $\phi_i$ by requiring that the E-fold number 
should be in the the range $50 < N < 60$ and then compute $n_s(\phi_i)$ and $r(\phi_i)$.  It is 
clear that the results depend on the chosen form of $V$: for example, the quadratic LFI model 
gives $n_s -1 \approx -2/N$ and $r \approx 8/N$, so, the relation $n_s -1 + r/4 = 0$ holds, which is in 
turn almost excluded by recent results of the PLANCK mission \cite{planck_1,planck_2,planck_3}.

\begin{figure}[ht] 
\begin{center} 
\epsfig{file=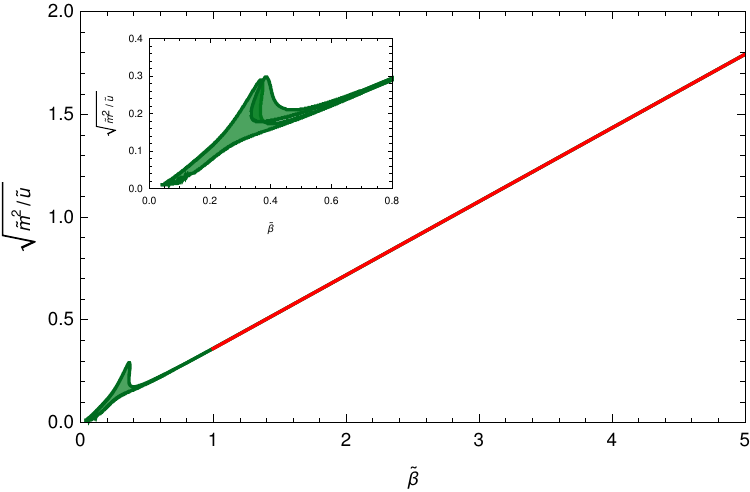,width=6.6cm}
\caption{\label{fig1}
Comparison of the Planck data with the slow-roll result of the 
MNI single-field inflationary potential (\ref{MNI}) with the specific 
choice for the constant term given in Eq.~\eqref{V2},
for $N = 55$. The best choice for the parameters is that located 
inside the green region. The red line corresponds to fitted line 
starting from the origin. In the inset we report the comparison for 
small $\beta$ of the best acceptance regions (for $V_2$ and $V_3$).} 
\end{center}
\end{figure}

We performed the described procedure for the potential (\ref{MNI}) with various values for the constant
term, with an ensuing comparison to CMBR data, see \app{mni_slow_roll}. 
The results are summarized in \fig{fig1}. The inset shows 
the best acceptance regions in the small $\tilde \beta$ limit (for $V_2$ and $V_3$); these may depend on 
the a particular choice of $V_0$. However, we have verified, by numerical calculations,
that in the large $\tilde \beta$ limit, each form of the MNI model considered by us gives the
same slow-roll result, i.e., the best acceptance region becomes a straight line which does not depend on 
the specific form of the MNI model chosen. This statement holds for the 
region $\tilde\beta \gtrsim 1.5$ and does not change when $N$ is varied between $50$ and $60$.
In general, because of the automatic retention of the minimum of the 
potential at zero, and because of the compact functional form, we prefer
the potential $V_2$, but we stress that our results are general.

Indeed, the straight line of \fig{fig1} does not depend on the 
particular choice for $V_0$ and a further free parameter of the model can be fixed by CMBR 
normalization. In addition, the most important fact that we evince from \fig{fig1} is that the 
(dimensionless) ratio $\tilde u \tilde \beta^2/\tilde m^2$ is also fixed. We denote it by $a$ and 
we observe that, of course, $\tilde u \tilde \beta^2/\tilde m^2=u \beta^2/m^2$. It follows that 
when one extracts the MNI parameters from a comparison with experimental PLANCK data, one 
does not actually have two free parameters, but just one. In other words, the dark green ``tail'' 
of the acceptance figure is a straight line, and the ratio $u \beta^2/m^2$ remains unchanged for 
large $\beta$. This points to a kind of universal behaviour in the large-frequency limit, i.e., at 
large $\beta$. In the following, we stick to this large-$\beta$ region, corresponding to small field 
inflation, leaving a detailed analysis of small-$\beta$ region for a future work. 

We conclude this Section by observing that the slow-roll study of the 
MNI potential has been performed preserving its sinusoidal functional form. 
One could, in principle, truncate the Taylor expansion of the 
MNI potential keeping only the constant, quadratic and quartic terms. 
One may find good agreement with the Planck data, 
but it would require a large explicit mass. However, 
such a large explicit (dimensionful) mass cannot be 
scaled down to the required value for the Higgs mass 
at the electroweak scale. We shall come back on this point in Section 
\ref{es_sec}.

%
% Fig 1
%
%

%------------------------------------------------------------------------------------
\subsection{Unification of Scales}
\label{gut_sec}
%------------------------------------------------------------------------------------

Based on the slow-roll analysis discussed above, one can fix almost all parameters of the MNI
(\ref{MNI}) potential, and there is only a single free parameter left. In this subsection, we show 
how this remaining free parameter can be determined by requiring a kind of unification of all 
scales at the Grand Unification Theory (GUT) scale $k = 2 \times 10^{16}$ GeV. 

The relation $u \beta^2/m^2 >1$ holds. The dimensionless ratio $a = u \beta^2/m^2$ is found to 
have the following value
\bea
a = \frac{\tilde u \tilde \beta^2}{\tilde m^2} \approx \left(\frac{0.70}{0.26}\right)^2 \approx 7.24\,,
\label{slope}
\eea
as extracted from the data of \fig{fig1}. We emphasize that the result 
(\ref{slope}), valid at large $\beta$, do not depend on the constant term 
$V_0$, as it is highly desirable. 

One can choose a particular point of that straight line 
where one finds the explicit mass of the MNI model,  
\bea
\tilde \beta \approx 270 \hskip 0.5cm \to \hskip 0.5cm m \sim 7 \times 10^{15} \mr{GeV} \,,
\eea
in the the range of the cosmological scale, i.e., $k_{\mr{cosmo}} \sim 10^{15} - 10^{16}$GeV
and close to the GUT scale. In this case the dimensionless ratio $a$ again has the value 
$a \approx 7.24$. One cannot choose larger value for $\tilde\beta$ because then the explicit 
mass would be larger than the typical energy scale of the system, i.e., the cosmological scale. 
Thus, our choice is a maximum for $\tilde \beta$.

\begin{figure}[ht] 
\begin{center} 
\epsfig{file=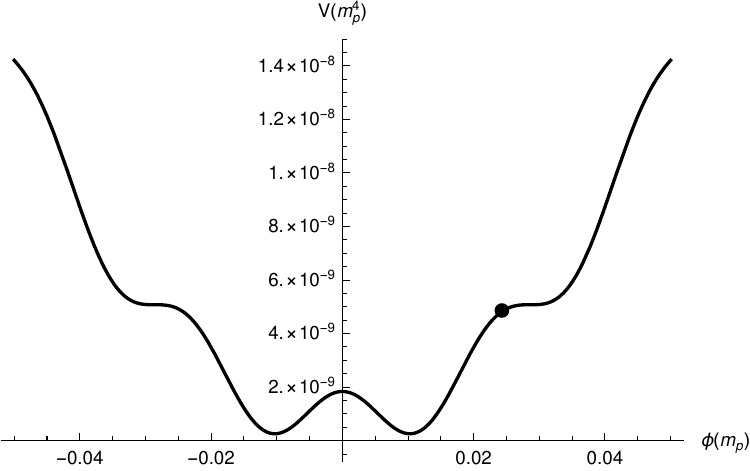,width=6.6cm}
\caption{\label{fig2}
The MNI potential (denoted in the text as $V_2$) at the GUT scale. 
The black blob denotes the initial value of the VeV in the inflationary period.} 
\end{center}
\end{figure}

With this particular choice $\tilde \beta \approx 270$,  the initial value for the VeV of the field is 
around $0.02 \,\, m_p$ (see \fig{fig2}).
%
% Fig 2
%
Thus,  we would like to emphasize that
the MNI potential~(\ref{MNI}) requires no large field values in the very beginning 
of the cosmic inflation, and so, it is more reliably supported by GUT motivated 
model building rather than typical large-field inflationary models, where one has to assume 
very large initial value for the VeV of the field ($4 - 20 \,\, m_p$). In addition, the initial value for 
the VeV of the MNI model  is again in the range of the GUT scale. Therefore, 
by the choice $\tilde \beta \approx 270$ one is somehow able to unify all scales. 

Finally, one should mention the following. If one would like to take into account all consequences 
of the quantum fluctuations of the inflaton field over inflation (where the rolling down of the VeV 
is basically considered as a classical process), one has to consider the RG 
running where the running scale is associated to the slowly rolling field itself \cite{lyth}.
This can be done in a reliable manner if the initial value for the VeV is around the momentum 
scale (which otherwise has been used as the scale for RG running) typical for inflation which is
the GUT scale (or cosmological scale). The MNI model~(\ref{MNI}) with the dimensionless 
frequency $\tilde \beta \approx 270$ fulfills the required criteria.

%------------------------------------------------------------------------------------ 
\section{Electroweak Scale}
\label{es_sec}
%------------------------------------------------------------------------------------

There is a strong interest to find a link between these scalar fields of the Higgs and inflationary physics 
\cite{higgs_inflation_2_1,higgs_inflation_2_2,higgs_inflation_2_3,higgs_inflation_2_4,higgs_inflation_1_2,%
higgs_inflation_1_1,higgs_inflation_1_3,higgs_inflation_1_4,higgs_inflation_1_7,higgs_inflation_1_6,higgs_inflation_1_5}. 
The Standard Model (SM) 
Higgs field is an SU(2) complex scalar doublet with four real components, and the underlying symmetry of the 
electroweak sector is $SU(2)_L \times U(1)_Y$, thus, the SM Higgs Lagrangian reads as
\begin{equation}
\label{sm_higgs}
{\cal L} = (D_\mu \phi)^\star (D^\mu \phi) - V(\phi) - \frac{1}{2} \Tr \, ({F}_{\mu\nu} {F}^{\mu\nu}) 
\end{equation}
with 
\begin{equation}
V = \mu^2 \phi^\star \phi + \lambda (\phi^\star \phi)^2
\end{equation}
and 
\begin{equation}
D_\mu = \partial_\mu + i g {\bf T} \cdot {\bf W}_\mu +i g' y_j B_\mu, 
\end{equation}
where the vacuum expectation of the Higgs field is either at zero field for $\mu^2>0$ or at 
$\sqrt{\phi^\star \phi} = \sqrt{-\mu^2/(2\lambda)} = v/\sqrt{2}$ for $\mu^2<0$ with $v = 246$ GeV known 
from low-energy experiments. The field can be parametrized around its ground state, where the unitary 
phase can be dropped by choosing an appropriate gauge.  
As a consequence of the Brout-Englert-Higgs mechanism \cite{englert_brout,higgs}, three degrees of freedom 
of the Higgs scalar field (out of the four) mix with weak gauge bosons. The remaining degree of freedom 
becomes the Higgs boson discovered at CERN's Large Hadron Collider \cite{ATLAS,CMS}. The complete 
Lagrangian for the Higgs sector of the SM with the single real scalar field $h$ reads
\begin{multline}
\label{higgs}
{\cal L} = \hf \partial_\mu h \partial^\mu h - \hf M^2_h h^2 - \frac{M^2_h}{2v} h^3 - \frac{M^2_h}{8v^2} h^4 \\
+ \left(M^2_W W^+_\mu W^{-\,\mu} + \hf M^2_Z Z_\mu Z^\mu \right) \left(1+2\frac{h}{v}+\frac{h^2}{v^2}\right),
\end{multline}
where $M_h = \sqrt{-2\mu^2} = \sqrt{2\lambda v^2}$. The measured value for the Higgs mass 
$M_h = 125.6 \, \mr{GeV}$ implies $\lambda = 0.13$. Incidentally, we note that the latter value is close to 
the predicted value based on an assumption of the absence of new physics between the Fermi and Planck 
scales and the asymptotic safety of gravity \cite{shapo_wett}. 

Extrapolating the SM of particle physics up to very high energies leads to an interpretation of the Higgs boson 
as the inflaton. Therefore, the most ``economical'' choice would be to use the same scalar field for Higgs and 
inflationary physics. The action can be defined either in the Jordan frame in which some function of the scalar 
field multiplies the Ricci scalar $R$, or in the Einstein frame in which the Ricci scalar is not multiplied by a 
scalar field~\cite{jordan_einstein}. To perform the slow-roll study, the action is usually rewritten in the Einstein 
frame and it takes the form for the case of minimal coupling to gravity,
\bea
\label{einstein}
&S = \int d^4x \sqrt{-g} \left[\dfrac{m_p^2 R}{2} - \hf g^{\mu\nu} \, \partial_\mu \phi \, \partial_\nu \phi - V(\phi) \right] \,,
\nn
&V \equiv \dfrac{\lambda}{4} \left(\phi^2 - v^2\right)^2 = \dfrac{M^2_h}{8 v^2} \left(\phi^2 - v^2\right)^2,
\eea
where the metric tensor being denoted by $g^{\mu\nu}$, $ \sqrt{-g}  \equiv  \sqrt{-{\rm det} \, g}$
while $\phi \equiv h$ and $V$ is the quartic-type double-well scalar potential of \eq{higgs},
\begin{equation}
\label{higgs_xi0}
V(\phi) = \frac{\lambda}{4} v^4 - \hf \lambda v^2 \phi^2 + \frac{\lambda}{4} \phi^4 \,.
\end{equation}
where the field variable is shifted as $h \to h+v$. The case of non-minimal inflation is discussed in \app{non_minimal_inf}.

Another proposal to build up the scalar sector is the Higgs inflation from false vacuum with minimal coupling 
to gravity where the SM Higgs potential is extended and assumed to develop a second (or more) minimum 
\cite{higgs_inflation_2_1,higgs_inflation_2_2,higgs_inflation_2_3,higgs_inflation_2_4}. The difficulty 
is to achieve an exit from the inflationary phase: one may introduce new fields, but than the attractive 
minimality of the model would be lost. 

Another possible drawback of the Higgs-inflaton potential to its applicability is that the measured 
Higgs mass is close to the lower limit, $126$ GeV, ensuring absolute vacuum stability within the SM 
\cite{NNLO_stability}. However, it was also shown \cite{bezrukov_rubio_shapo} that traditional Higgs 
inflation can be possible within a minimalistic framework even if the SM vacuum is not completely stable. 
Various polynomial Higgs potentials have been studied by functional RG \cite{higgs_frg_1,higgs_frg_2} 
and reported no stability problems. 

All versions of the MNI model studied by us
contain two adjustable parameters (the ratio $u/m^2$ and the 
frequency $\beta$) and a normalization (the field-independent terms has been fixed by us). The 
Taylor expansion of the MNI model (\ref{MNI}) recovers the SM Higgs potential 
\eq{higgs_xi0} up to quartic terms and the parameters can be related:
\begin{equation}
V_{\rm MNI} \approx V_0 + \hf (m^2 - u\beta^2) \phi^2 + \frac{1}{24} u \beta^4 \phi^4 + 
{\mathcal O}(\phi^6) \,,
\label{def_14}
\end{equation}
so that
\begin{equation}
\label{msg_expand}
\lambda v^2 \equiv (u\beta^2 - m^2), \hskip 0.4cm \lambda \equiv \frac{1}{6} u \beta^4. 
\end{equation}
Thus the MNI model (\ref{MNI}) can be considered as an UV completion 
of the SM Higgs potential. Further details of the UV completion is shown in \app{uv_completion}.
The measurable quantities are related to the parameters of the model 
according to the following relations: 
\beq
\label{msg_higgs_params}
M_h \equiv m \sqrt{2\left(\frac{u\beta^2}{m^2} - 1\right)}, \hskip 0.4cm v 
\equiv \frac{1}{\beta} \sqrt{\frac{6 (u\beta^2/m^2 - 1)}{u\beta^2/m^2}}.
\eeq
Their low-energy/IR values are given at the electroweak scale by
\beq
\label{IR_higgs}
M_{h,\mr{IR}} = 125 \,\, \mr{GeV}, \hskip 0.5cm v_{\mr{IR}} = 245 \,\, \mr{GeV} \,,
\eeq
at the scale $k_{\mr{IR}} \sim 250 {\mr{GeV}}$.

Notice that taking into account higher order terms in Eq. (\ref{def_14}) gives 
rise to results consistent with (\ref{msg_higgs_params}), as discussed in 
\app{effect_vev}.

Let us note, that the Higgs mass and VeV defined by \eq{msg_higgs_params} can be calculated 
also at the cosmological scales. For example, in the large $\tilde \beta$ region the slow-roll study 
produces values for the Higgs mass and VeV in the same order of magnitude which serves as a 
high-energy/UV scale 
\begin{subequations}
\begin{align}
\label{UV_higgs}
M_{h,\mr{UV}} \sim & \; v_{\mr{UV}} \sim 10^{15} \, \mr{GeV} \,, \\
m_{\mr{UV}} \sim & \; 10^{15} \mr{GeV}, \hskip 0.2cm \beta_{\mr{UV}} \sim 10^{-15} \, \mr{GeV}^{-1} \,,
\end{align}
\end{subequations}
at the scale $k_{\mr{UV}} \sim 10^{15} {\mr{GeV}}$.
These need to be scaled down (by orders of magnitude) to their measured values at the electroweak scale 
\eq{IR_higgs}. The integration of high-energy modes is a well-known method which provides us this scaling 
down by the successful integration of the field fluctuations. Therefore, let us discuss the integration of the 
high-energy modes in the MNI model in the post-inflation period.

%------------------------------------------------------------------------------------
\section{Mode Integration}
\label{mi_sec}
%------------------------------------------------------------------------------------ 

In principle, a comprehensive study of the integration of high-energy modes requires an accurate treatment 
of gauge and fermion fields and not just a single scalar potential. However, the MNI model has a very 
important feature, namely it contains a periodic and a quadratic self-interaction term (apart from the trivial
constant term). It allows to perform explicitly, in an easy way, the integration of modes up to the electroweak scale. 
The explanation is the following. The integration of modes is done by solving the RG flow differential equation 
where not the action but its Hessian (second derivative) appears on the one side \cite{eea_rg_1,eea_rg_2}.
So, even if the scalar field couples to any gauge or fermion fields in the action (which are not higher than 
quadratic in the scalar field), the Hessian remains periodic in the scalar sector. Therefore, it represents a 
reliable approximation if one looks for the solution of the equation among scalar periodic functions and neglect 
the effect of other fields.

%--------------------------------
\subsection{%Nonperturbative Renormalization Group
Flow Equation for the Potential}
%--------------------------------

A field theory closely related to the MNI model (\ref{MNI}), namely, 
the massive sine-Gordon scalar model, 
was extensively studied in $d=2$ dimensions by functional RG method \cite{msg_lpa_1,msg_lpa_2,msg_lpa_3,msg_beyond_lpa}. 
Here, we consider $d=4$ dimensions. The procedure, standard in RG approaches, 
consists in eliminating the high-energy modes by integrating them. This gives an equation for the flow of the potential, and the general fact is that this 
equation is a functional one. When the high-energy modes are integrated out, 
the potential therefore 
becomes scale-dependent, $V \to V_k$, 
where with the symbol $V_k$ we denote that the modes with momenta larger than 
$k$ have been integrated out. To determine the dependence on the scale $k$ 
one has to solve the RG 
flow equation which has the following form at the level of the 
local potential approximation where higher-order derivatives of the field are 
neglected \cite{eea_rg_1,eea_rg_2},
\bea
\label{lpa_erg_pre}
k \partial_k {V_k(\varphi)} =  \hf \int \frac{d^d p}{(2\pi)^2} \, \frac{k\partial_k R_k}{R_k + p^2 + \partial^2_{{\phi}} V_k} \,,
\eea 
where $R_k = R_k(p^2)$ is a cut-off function, 
which in the functional RG jargon is called regulator function. Then,
one can turn to dimensionless quantities, denoted by a tilde superscript as 
written in Eqs.~(\ref{dimless_param}).  
The RG flow equation becomes
\bea
\label{lpa_erg_dimless}
\left(d-\frac{d-2}{2} \tilde \varphi \partial_{\tilde\varphi} + k \partial_k \right) \tilde V_k(\tilde\varphi) = \hskip 2.0cm \nn
- \alpha_d  \int_{0}^{\infty}  dy  \, \frac{r' \, \, y^{\frac{d}{2}+1}}{[1+r] \, y \, + \partial^2_{\tilde{\phi}} \tilde{V}_k} \,,
\eea 
with $\alpha_d = \Omega_d/(2(2\pi)^d)$ and 
$\Omega_d = 2 \pi^{d/2}/\Gamma(d/2)$, where it is intended we have to put 
$d=4$. Moreover $r(y)$ is the dimensionless regulator defined by 
$r(y) = R_k(p^2)/p^2$, with $y=p^2/k^2$ and $r' = dr/dy$.
The integral in the right-hand-side of Eq.~\eq{lpa_erg_dimless} 
can be performed analytically by 
the appropriate choice of the regulator function, for example by using 
%the optimized regulator 
$r(y) = (1/y -1) \Theta(1-y)$. The results should be independent on the 
particular regulator which has been chosen: this issue has been discussed 
in considerable detail in the functional RG literature, so we just pass 
to present our results. 

\begin{figure}[ht] 
\begin{center} 
\epsfig{file=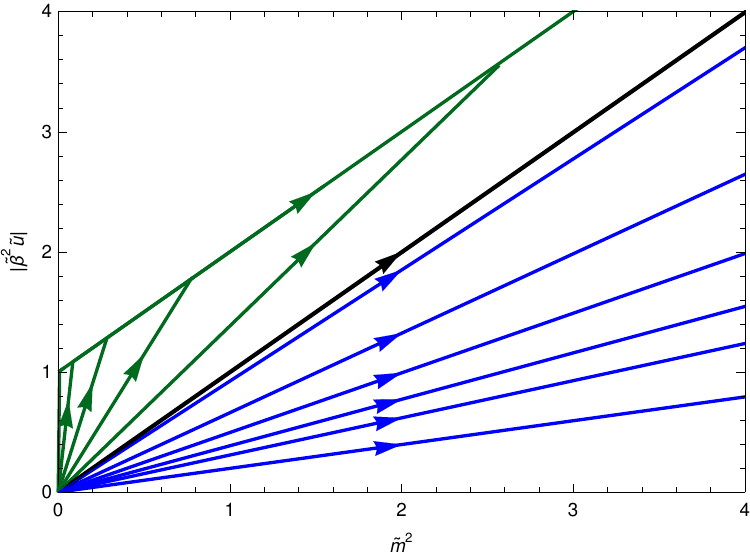,width=6.4cm}
\caption{\label{fig3}
Flow of the MNI model showing two phases separated by a black line with a unit slope.
The blue lines corresponds to the symmetric phase, while the green lines correspond to the SSB phase.
} 
\end{center}
\end{figure}

\begin{figure}[ht] 
\begin{center} 
\epsfig{file=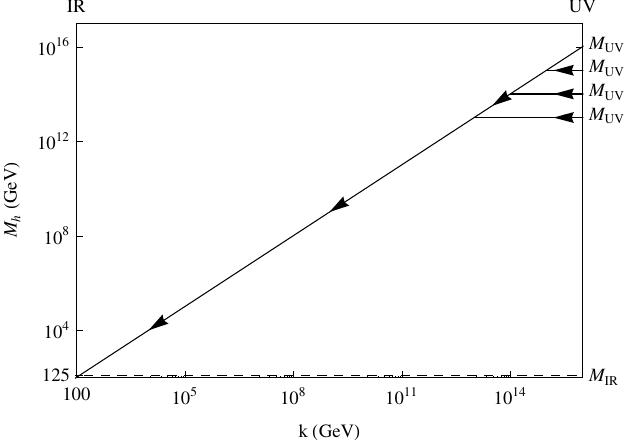,width=6.6cm}
\caption{\label{fig4}
Flow of the Higgs mass from the cosmological (UV) scale to the electroweak (IR) scale 
obtained from the study of the MNI model. The trajectories merge into a single line showing UV insensitivity.
} 
\end{center}
\end{figure}

%
%,  one obtains for the dimensionless scaling potential $\tilde{V}_k$ the following 
%RG equation,
%%
%\beq
%\label{opt}
%\left(d- \frac{d-2}{2} \tilde{\phi} \partial_{\tilde{\phi}} + k\partial_k \righ%t) \tilde{V}_k=
%\frac{2 \alpha_d}{d} \, \frac{1}{1+\partial^2_{\tilde{\phi}} \tilde{V}_k} \,,
%\eeq
%%
%which has to be taken in $d=4$. One has to look for the solution of 
%Eq.~\eq{opt} in the functional form of the MNI model. 
Two comments are anyway in order: {\em i)} the mode integration procedure 
above described give the same results for all variants of the MNI model 
($V_1$, $V_2$ and $V_3$) since the derivation with respect to the field 
eliminates any dependence on field-independent terms; {\em ii)} the results 
presented in \fig{fig3} and \fig{fig4}, and the UV-insensitivity property, 
do not depend on the choice of the 
regulator.

%One should separate the periodic and non-periodic parts where the latter results in a trivial scaling for %
%the dimensionless frequency $\tilde{\beta}_k^2$ and for the dimensionless mass %$\tilde{m}_k^2$, while the 
%corresponding dimensionful couplings $\beta^2$ and $m^2$ remain constant over the flow. Solving the 
%periodic part of Eq. (\ref{opt}), 

Our results can be summarized by observing that one 
finds two phases controlled by the dimensionless quantity 
$\tilde{u}_k \tilde{\beta}_k^2/\tilde{m}_k^2$,  which tends to a constant in the IR limit. In the ($Z_2$) 
symmetric phase the magnitude of this constant is arbitrary (and depends on the initial conditions), 
but always smaller than one, i.e., $\lim_{k\to 0} \vert \tilde{u}_k \tilde{\beta}_k^2/\tilde{m}_k^2 \vert <1$
(see the blue lines of \fig{fig3}). 

In the spontaneously broken (SSB) phase, the IR value of the magnitude of the ratio is exactly one 
(independently of the initial values) which serves as an upper bound, see green lines of \fig{fig3}. 
The black line separates the two phases. 
%
% Fig 3
%

In other words, trajectories in the SSB phase (green lines) merge into a master trajectory (green line 
parallel to the black one) of \fig{fig3} which implies 
\beq
\bar{u}_k \tilde{\beta}_k^2  = 1 + \tilde{m}_k^2 \,,
\eeq
and it results in the following scaling:
\beq
\label{scaling}
\frac{\bar{u}_k \tilde{\beta}_k^2}{\tilde{m}_k^2} -1 = \frac{1}{\tilde{m}_k^2} =  \frac{k^2}{m^2_{\mr{UV}}}.
\eeq
This scaling relation is valid when the running is determined by the master trajectory which, apart from the very 
beginning of the running, is always the case in the SSB phase. We observe that 
Eq. (\ref{scaling}) has been obtained with a non-perturbative approach 
for the MNI potential (\ref{MNI}), but one may expect that it 
is valid for other classed of inflationary potentials, as we are going to 
discuss in Section \ref{comments} and \ref{concl}. However, the MNI 
potential (\ref{MNI}) is well suited to find Eq. (\ref{scaling}) 
in a particularly transparent way.

%Under these assumptions,
Eq.~(\ref{scaling}) can be used to determine 
the IR value of the Higgs mass and VeV from the UV initial conditions (\ref{UV_higgs}) and compared to 
the known results of (\ref{IR_higgs}). Indeed by substituting (\ref{scaling}) into (\ref{msg_higgs_params}) and 
assuming running parameters $\tilde{u}_k \tilde{\beta}_k^2/\tilde{m}_k^2 = u_k \beta^2/m^2$, one gets
\begin{equation}
M_{h}(k) = 
m_{\mr{UV}} \sqrt{2\left(\frac{\tilde{u}_k \tilde{\beta}_k^2}{\tilde{m}_k^2} -1\right)}. 
\end{equation}
Now, one uses Eq. (\ref{scaling}) and obtains
\begin{equation}
M_h(k)= m_{\mr{UV}} \, \sqrt{2} \, \sqrt{\frac{k^2}{m_{\mr UV}^2}}= \sqrt{2} 
\,  \cancel{m_{\mr{UV}}} \frac{k}{\cancel{m_{\mr{UV}}}} = \sqrt{2} k, 
\end{equation}
where the cancellation of the UV mass has been made 
evident. We observe, for absolute clarity, that this property does not 
depend of the specific form of the flow (\ref{lpa_erg_dimless}), 
but is a consequence of the fact that the ratio $u \beta^2/m^2$ 
is constant. 

Similarly, one gets
\begin{equation}
v(k) = %\frac{1}{\beta} \sqrt{\frac{6 (u_k\beta^2/m^2 - 1)}{u_k\beta^2/m^2}} 
\frac{1}{\beta_{\mr{UV}}} \sqrt{\frac{6 (\tilde{u}_k 
\tilde{\beta}_k^2/\tilde{m}_k^2-1)}{\tilde{u}_k \tilde{\beta}_k^2/\tilde{m}_k^2}} 
=  \frac{\sqrt{6} k}{m_{\mr{UV}} \beta_{\mr{UV}}}  \,,
\end{equation}
which has important consequences. Since $k_{\mr{IR}}$~=~250~GeV,  it provides the required IR values \eq{IR_higgs}, 
at least the same order, in accordance with measurements. Furthermore, the IR values for the Higgs mass becomes 
independent of the UV initial parameter (see \fig{fig4}). 

We conclude this Section by observing  
that the (dimensionful) Higgs mass has a weak RG running in 
the framework of perturbative RG in the SM \cite{Peskin}. 
We are in agreement 
with this fact, since in our model the (dimensionful) 
mass $m$ has no running at all. However, with the non-perturbative approach 
followed in the present paper, 
the Higgs mass is an effective mass and it is not the explicit mass $m$. 
This is shown in \fig{fig4} clearly demonstrating the linear dependence of the 
effective Higgs mass obtained using the MNI model (\ref{MNI}). 
%
% Fig 4
%
%

%------------------------------------------------------------------------------------
\subsection{Comments on Universality}
\label{comments}
%------------------------------------------------------------------------------------

The previous derivation show that the details at cosmological scale do not influence the quantities at the 
electroweak scale. This property manifests itself through the cancellation of the parameter $m$ of the MNI
determined at cosmological scale using PLANCK data. In practice, whatever is the value of $m$ and $\beta$ 
at cosmological scale, one gets the same value for the Higgs mass, which depends only on the chosen energy 
scale at electroweak, IR scale, which we can denote by $k_{\rm ew}$. This is evident from the universality of 
the flow, as depicted in \fig{fig4}, where all trajectories flow into the same point.  Our analysis also shows that 
the initial condition, which we may denote by $k_{\rm start}$, chosen as starting point for the mode integration, 
also does not play a role.

Now, the question that naturally emerges is the following: 
is this property specific to the MNI alone or not? Before answering, we pause to comment that 
in general, the MNI is very suited to show the property of independence of the 
Higgs mass from the initial conditions and from the values of the parameters 
at cosmological scale, since the mode integration can be performed 
straightforwardly using known results from field theory. Actually, the 
cancellation of the mass $m$ emerge naturally and with a minimally simple 
calculation. However, this usefulness of the MNI does not imply that 
it is the only model exhibiting such features. Indeed, we expect that other 
two-parameter inflationary model with $Z_2$ symmetry will give the same results. 
The reason for such expectation is three-fold: from one side, during the mode 
integration, all powers of the field are generated and the interplay of 
the minima depends on these powers. Such a generation is not specific of the 
MNI model and will be exhibited by other models: actually, we can say the 
opposite, namely that if the model has an Higgs mass depending 
on the parameters determined from cosmological data, then most probably 
it could be characterized as unphysical. So, this led to conclude that extending an inflationary 
model towards low energies should produce the same Higgs mass provided 
that $k_{ew}$ is fixed, and the MNI is in our opinion just a simple model 
showing this result. From the other side, it is well known that 
mode integration makes the potential tend to a convex form 
starting from a concave one, and this is in agreement with 
$M_h(k_{ew})/m \ll 1$. Finally, it should be noted that for scalar theories 
in four dimensions, the only relevant and marginal couplings are the quadratic 
and the quartic one respectively. When considering more 
complicated functional forms for the potential,
the scaling arguments cannot be straightforwardly 
applied, and the irrelevance of higher-order field terms is not evident from the $\beta$-functions. 
Even in this case, the simple scaling arguments are expected to hold, and so we expect 
the irrelevance from the UV conditions to hold even for different UV completions.

%------------------------------------------------------------------------------------ 
\section{Conclusions and Outlook}
\label{concl}
%------------------------------------------------------------------------------------ 

%
A good candidate for an inflationary potential should fulfill the following conditions,
{\it (i)} have an as-simple-possible functional form (with the smallest possible number of parameters), 
{\it (ii)} provide the best agreement with observations, 
{\it (iii)} be as ``economical" as possible in term of the formulation of the theory. 
In this paper, we propose the extended version of the massive sine-Gordon theory
as an inflationary potential and we refer to it as the massive Natural Inflation (MNI) model.
We show that adding the mass term to the periodic potential 
produces a remarkably improved agreement with the Planck results. We attribute such improvement to the fact that 
it has infinitely many minima, but they are 
non-degenerate and with tunable controllable energy difference, 
providing a way to be as much as possible both ``$\phi^2$ and not-$\phi^2$''. The crucial point emerging from a 
careful analysis of different potentials is that the inflationary potential should have a concave region, and the mass 
term in the MNI potential add such an overall convexity in presence of the many minima.

To explore the issue of the convexity, we used the known 
renormalization group (RG) results for the MNI model, extending them 
to $d=4$, to perform explicitly the mode integration. We then determined the  
phase diagram associated to the running determined by the 
slow-roll conditions having in mind the post-inflation period. The obtained values for the ratio 
$\vert \tilde u_k \tilde \beta_k^2/\tilde m_k^2\vert$ 
are found in the phase of spontaneous symmetry breaking (above the critical line). Thus, the study 
of the MNI model shows that slow-roll conditions represent strong constraints on the RG running i.e., it stays in 
its broken phase. Moreover, the mode integration produces a convex (dimensionful) potential in agreement with the 
theoretical requirement of the convexity of the effective potential. 

In conclusion, we introduced a model for cosmological inflation, based on a UV completion of the Higgs potential, 
which has the following properties:
\begin{itemize}
\item it works at cosmological level, and its parameters can be fixed from PLANCK and BICEP2 experimental data. 
(This property is shared with many other models which are consistent with available data at the 
cosmological scale, but still constitutes an important consistency check of our model.)
\item Our model also works at the electroweak scale which means its parameter can be fixed  by comparison
with experimental data. This property, by contrast, is not shared by other UV-admissible models 
which work at the cosmological scale.
Within our assumptions, the Higgs mass is found to be independent of the parameters 
found at the cosmological scale.
\item We argue that these properties are shared by other two-parameter inflationary models, but our model provides 
an ideal framework to explicitly show it in a straightforward way.
\end{itemize}
We conclude that the model is valid both at cosmological and electroweak scales.

Finally, we comment on the consequences of our results related to three major issues of inflationary cosmology 
which have and are intensely discussed. The first concerns the identification of the inflaton with the Higgs field, 
i.e., whether the inflationary potential can be be extended to electroweak scale in a reliable way. The second is 
whether the high-energy properties of the theory affect or not the low-energy ones: when there is not such influence, 
one talks about UV-insensitivity \cite{fumagalli}. The third issue is related to the fact that many inflationary models can 
be and have been proposed \cite{encyc}, so that it may be asked whether and how the correct model should be chosen. 
The fact that one has to choose an {\em ad hoc} potential with a fine tuning of its parameters and of the initial conditions 
is certainly an argument against inflationary scenarios. Our paper brings a contribution to this on-going discussion, since 
our results suggest that the answer to first question is ``yes", in the sense that the results for the Higgs mass does not 
depend on the values of the model parameter at cosmological scale and that one can explicitly build a model valid at 
cosmological and electroweak scale. The answer to second question, looking at the results of the MNI analysis presented 
in the present paper, is that there is UV-insensitivity, as the MNI clearly shows with the cancelation of the mass $m$ at 
cosmological scale in determining $M_h(k_{ew})$. For the third question we conclude ``no", in the sense that other models 
with non-fine tuned initial conditions share the same properties. The affirmative answer to the first question supports
Higgs inflationary models. If one were to insist that a single specific model,
with uniquely determined parameters, should explain the physical 
properties at all scales, then the model-independence and UV-insensitivity could be 
considered as an argument against Higgs inflation. 
However, if one is taking an approach like the renormalization group approach {\em \`{a} la} Wilson, then the 
model-independence (and UV-insensitivity) 
is not only desirable, but, in a sense, required. 

Summing all the radiative corrections to the quartic 
Higgs potential (and of course having at disposal 
more experimental data) could shed light on this apparent contradiction 
and on the form of the 
effective action at the cosmological scale. 
However, we think that the top-down 
approach based on a guessed action at high-energies may give very useful informations to clarify the properties 
that the UV action should have in order to fulfill the requested requirements at all scales. This approach is opposite 
to the more standard one bottom-up, which starts from the 
low-energy properties and tries to obtain the action at 
high-energies, and it may give complementary information. 
This is one of the reasons for which we hope that the results presented 
in this paper may contribute to the discussion on the validity of 
specific inflationary models.

In conclusion, we found a class of MNI models which show UV insensitivity and can act as viable candidates for
an inflaton mechanism, via integration of the high-energy modes. While it will be impossible, from low-energy 
experiments alone, to determine which is the correct one, one could confirm that only rather moderate extensions 
of the Standard Model (in this case, the addition of periodic terms) would be required for a consistent extrapolation 
to very high energy scales, the latter being of cosmological relevance.

{\it Acknowledgements.---} 
This work was supported by the J\'anos Bolyai Research Scholarship of the Hungarian Academy of Sciences and
by the \'UNKP-17-3 New National Excellence Program of the Ministry of Human Capacities. Useful discussions with 
F. Bianchini, G. Gori, J. Rubio, Z. Trocsanyi and G. Somogyi are gratefully acknowledged. 
Support from the National Science Foundation (Grant PHY--1710856) also is being acknowledged.
Financial support by the CNR/MTA Italy--Hungary 2019--2021 Joint Project "Strongly interacting systems in confined 
geometries" is gratefully acknowledged.

\begin{appendices}

%------------------------------------------------------------------------------------ 
\section{Inflationary predictions}
\label{mni_slow_roll}
\begin{figure}[ht] 
\begin{center} 
\epsfig{file=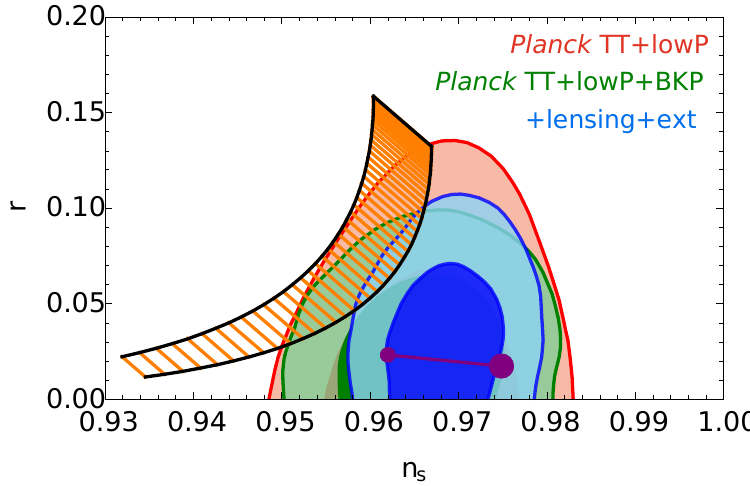,width=7.0cm}
\caption{\label{fig5}
CMBR parameters, i.e., scalar tilt $n_s$ and tensor-to-scalar ratio $r$ derived {\em (i)} for the 
Natural Inflation model for various frequencies  (orange line segments) and {\em (ii)} for the MNI model 
with fixed ratio $\tilde u/\tilde m^2 \sim 1/(0.22)^2$ and frequency $\tilde\beta \sim 0.3$ (purple line segment).
Both are compared to results of the Planck mission \cite{planck_1,planck_2,planck_3} where dark color regions 
stand for 95\% CL and light color regions correspond to 68\% CL.
} 
\end{center}
\end{figure}

%------------------------------------------------------------------------------------ 
In this appendix we discuss the applicability of the Natural Inflation (NI) and the Massive Natural Inflation (MNI) 
potentials for cosmology. The slow-roll parameters ($\epsilon$, $\eta$, $n_s$, $r$) of the NI model reads as
\begin{align}
\epsilon =& \frac{\tilde\beta^2}{2} \cot^2\left(\frac{ \tilde\beta \tilde\phi}{2} \right), \hskip 0.4cm
\eta = \frac{\tilde\beta^2}{2} \frac{\cos(\tilde\beta \tilde\phi)}{\sin^2 \left( \frac{ \tilde\beta \tilde\phi}{2} \right)},
\nn
N =& -\frac{2}{\tilde\beta^2}  \log \cos\left(\frac{\tilde\beta \tilde\phi}{2}\right) \Big|^{\tilde\phi_i}_{\tilde\phi_f}
\nn
n_s-1& \approx \tilde\beta^2 \left[1-2 \sin^{-2} \left(\frac{\tilde\beta \tilde\phi}{2}\right) \right], \,\,\,\,\,\,
r \approx 8 \tilde\beta^2 \cot\left(\frac{\tilde\beta \tilde\phi}{2}\right)
\nonumber
\end{align}
which imply the relation $n_s-1+\frac{r}{4}=-\tilde\beta^2$ which is very similar to that of obtained for the 
quadratic monomial potential but having a dependence on the frequency $\tilde\beta$, thus, it appears as 
an additional parameter which can be tuned ($\tilde \beta \sim 0.15$) to achieve a better agreement with 
the Planck data, see orange line segments of \fig{fig5}.

%
% Fig 5
%

The slow-roll parameters of the MNI model \eq{V1},
\begin{align}
\epsilon=& \hf \left( \frac{\frac{\tilde u}{\tilde m^2} \tilde\beta \sin(\tilde\beta \tilde\phi)+ \tilde\phi}{ \frac{\tilde u}{\tilde m^2} 
\left[1- \cos(\tilde\beta \tilde\phi)\right]+ \hf \tilde\phi^2 }  \right)^2
\\
\eta=& \frac{ \frac{\tilde u}{\tilde m^2} \tilde \beta^2 \cos(\tilde \beta \tilde \phi)+1}{ \frac{\tilde u}{\tilde m^2} \left[1- \cos(\tilde\beta \tilde\phi)\right]+ \hf \tilde\phi^2 }
\\
N=& -\int_{\tilde\phi_i}^{\tilde\phi_f} d\tilde\phi \frac{ \frac{\tilde u}{\tilde m^2} 
\left[1- \cos(\tilde\beta \tilde\phi)\right]+ \hf \tilde\phi^2}{\frac{\tilde u}{\tilde m^2} \tilde\beta \sin(\tilde\beta \tilde\phi) + \tilde\phi}
\end{align}
depend on the ratio $\tilde u/\tilde m^2$ and the frequency $\tilde\beta$. The inclusion of the explicit mass 
term ($\hf m^2 \phi^2$) in the periodic (i.e., NI) potential improves the inflationary predictions of the model, see \fig{fig5}.

%------------------------------------------------------------------------------------ 
\section{Non-minimal Inflation}
\label{non_minimal_inf}
%------------------------------------------------------------------------------------ 
In this appendix we discuss the case of a large non-minimal coupling to gravity \cite{higgs_inflation_1_1,higgs_inflation_1_2,higgs_inflation_1_3,%
higgs_inflation_1_4,higgs_inflation_1_5,higgs_inflation_1_6,higgs_inflation_1_7}, where the interpretation of the Higgs boson
as the inflaton results in the following action in the Jordan frame 
\begin{eqnarray}
\label{jordan_frame}
&S = \int d^4x \sqrt{-\bar{g}} \frac{m_p^2}{2} \left[ F(\tilde h) \bar{R} -
\bar{g}^{\mu\nu} \partial_\mu \tilde h \partial_\nu \tilde h - 2U(\tilde h) \right] \,,
\nn
&F(h) = 1+\xi \tilde h^2, \hskip 0.2cm  U(\tilde h) = 
m_p^2 \dfrac{\lambda}{4} \left(\tilde h^2 - \dfrac{v^2}{m^2_p}\right)^2 \,,
\end{eqnarray}
where $\tilde h$ is the dimensionless Higgs scalar ($h = m_p \tilde h$), $\xi$ is a new parameter and 
$\bar{g}^{\mu\nu}$ is the metric in the Jordan frame. Of course, by convention,
$ \sqrt{-\bar{g}}  \equiv  \sqrt{-{\rm det} \, \bar{g}}$, while $U(\tilde h)$ is the dimensionless quartic-type 
double-well scalar potential with a $Z_2$ symmetry, equivalent to the SM Higgs potential of 
(\ref{higgs}). In order to show this one has to {\it (i)} introduce a dimensionful field variable, 
{\it (ii)} replace the quartic self-coupling by the Higgs mass, i.e., $\lambda = M^2_h/(2 v^2)$ as is 
indicated below Eq.~\eq{higgs}, and {\it (iii)} shift the field variable as $h \to h+v$.  

To perform the slow-roll study, the action is usually rewritten in the Einstein frame and it takes the form 
\bea
\label{einstein_frame}
&S = \int d^4x \sqrt{-g} \left[\dfrac{m_p^2 R}{2} - \hf g^{\mu\nu} \, \partial_\mu \phi \, \partial_\nu \phi - V(\phi) \right] \,,
\nn
&V \equiv m_p^2 \dfrac{U}{F^2}, \hskip 0.2cm \dfrac{d\phi}{d\tilde h} = m_p 
\dfrac{\sqrt{1 + \xi(1 + 6\xi)\tilde h^2}}{1+\xi \tilde h^2} \,,
\eea
where the metric tensor being denoted by $g^{\mu\nu}$. For $\xi \neq 0$, the Higgs-$\xi$ inflaton potential reads
\beq
V(\phi) = \frac{m_p^4 \lambda}{4 \xi^2} \left[1 - \exp\left( -\sqrt{\frac23} \dfrac{\phi}{m_p} \right) \right]^2,
\eeq
which is considered as a zero parameter model since the overall factor of the potential is entirely 
determined by the amplitude of the CMBR anisotropies. In the framework of this inflationary scenario, 
one needs to take extra precautions in order not to violate perturbative unitarity.

%------------------------------------------------------------------------------------ 
\section{UV completion}
\label{uv_completion}
%------------------------------------------------------------------------------------ 
Alternatively, the MNI-type UV completion of the SM Higgs potential can be formulated as 
\beq
V = \hf M^2 \phi^\star \phi + u[\cos(b \sqrt{\phi^\star \phi})-1] \,,
\eeq
which recovers the scalar potential in \eq{sm_higgs} after its Taylor expansion with 
$\mu^2 \equiv 1/2 \,\, (M^2 - u b^2)$ and $\lambda \equiv  \tfrac{1}{24} \,\, u b^4$. Performing the same
parametrization of the field used in order to get from \eq{sm_higgs} to \eq{higgs} one finds
\beq
V = \hf \,\, \frac{M^2}{2} (h + v)^2 + u\left[\cos\left(\frac{b}{\sqrt{2}}(h + v)\right)-1\right],
\eeq
which is identical (apart from constant terms) to the MNI model \eq{MNI} 
after shifting the field by a constant $h+v \to \phi$ and introducing $m^2 \equiv M^2/2$ and
$\beta \equiv b/\sqrt{2}$. This validates that the MNI model is a suitable extension of the SM Higgs
potential. The broken symmetric case corresponds to parameters where $1<u\beta^2/m^2 = ub^2/M^2$, 
i.e., where $\mu^2 <0$.

%------------------------------------------------------------------------------------ 
\section{Effect of higher order terms on the vacuum expectation value}
\label{effect_vev}
%------------------------------------------------------------------------------------ 

\begin{figure}[ht] 
\begin{center} 
\epsfig{file=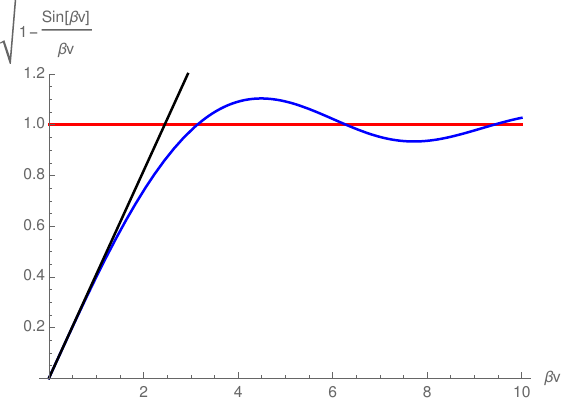,width=6.6cm}
\caption{\label{vev}
The first order approximation of the right hand side of 
(\ref{eq_for_vev_from_Vmin}) (black line). T
he red line is the maximum value of the left hand side of 
(\ref{eq_for_vev_from_Vmin}) while the blue curve is the exact result.}
\end{center}
\end{figure}

The vacuum expectation value (VEV) $v$ is defined by the value of the 
field ($\phi$) at the minima of the potential:
\beq
V(\phi=v)=\min(V).
\eeq
In our case the VEV is the closest solutions to zero of the equation
\beq
\left. \frac{dV}{d\phi}\right|_{\phi=v}=0 \,,
\eeq
which for the MNI model writes as
\beq
m^2 v -\beta u \sin(\beta v)=0 \,.
\eeq
One obtains
\beq
\label{eq_for_vev_from_Vmin}
\sqrt{\frac{u \beta^2/m^2-1}{u \beta^2/m^2} }=
\sqrt{1-\frac{\sin(\beta v)}{\beta v}} \,.
\eeq
The left hand side is smaller than one and we remind we are 
looking for the solutions for $v$ which is the closest to zero. 
Thus the calculations with only the first order of the right hand side 
(which is equivalent to keep only the $\phi^n$ terms, 
with $n \leq 4$, in the 
potential) is a reasonable approximation, see \fig{vev}. 

%
% Fig 1
%

We conclude that taking into account the higher terms changes the scaling 
of the VEV, which anyway stays very close to the first order 
approximation written in the paper.

\end{appendices}

\end{document}